\documentclass[
    ,final            
  ]
  {aipproc}

\layoutstyle{6x9}

\begin{document}

\title{Testing BSM physics through correlations between flavor observables}


\author{Monika Blanke}{
  address={Department of Physics, LEPP, Cornell University, 142 Sciences Drive, Ithaca, NY 14853, USA}
}

\begin{abstract}
I provide an overview of the connections between flavor violating observables in the $K$ and $B$ meson systems beyond the Standard Model (BSM). Model independent correlations, both in and beyond the MFV framework, as well as results obtained in specific models are discussed.
\end{abstract}

\maketitle


\section{Introduction}

With the LHC running and its experiments collecting data at an impressive rate, particle physics has entered a very exciting era. It will be a matter of years until we know whether the Standard Model (SM) provides an accurate description of electroweak symmetry breaking and the physics up to the TeV scale. If eventually new particles are discovered by Atlas and CMS, complementary information from the intensity frontier will be needed in order to understand their nature. If on the other hand no hints of new physics (NP) appear in the high-$p_T$ data, high-precision measurements will provide a powerful tool to probe particles too heavy to be directly accessible at LHC energies.

A particularly important class of high-precision low energy observables are rare decays of $K$ and $B$ mesons which probe flavor violating interactions and are highly sensitive to new sources of flavor violation beyond the CKM matrix. In addition specific correlations between flavor violating observables allow for a clear distinction of different NP scenarios.

\section{Flavor correlations in (C)MFV}

In order to account for the absence of large NP phenomena in the flavor sector, the Minimal Flavor Violation (MFV) hypothesis \cite{Buras:2000dm,D'Ambrosio:2002ex,Chivukula:1987py,Hall:1990ac} is a powerful ansatz. In this framework, all flavor violating effects are exclusively governed by the SM Yukawa couplings. Consequently, as all flavor violating transitions are then governed by the well-known quark masses and CKM parameters, specific correlations are predicted. In addition MFV provides an appealing alternative to R-parity in supersymmetric models \cite{Nikolidakis:2007fc,Csaki:2011ge} by effectively suppressing the dangerous R-parity violating interactions.

\begin{figure}
\begin{minipage}{7.3cm}
\includegraphics[width=.98\textwidth]{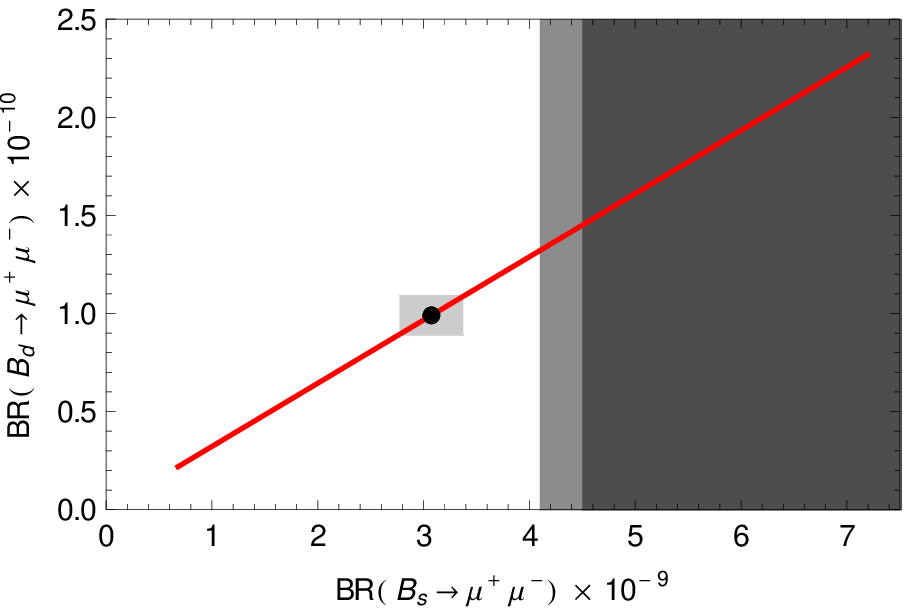}
\end{minipage}
\begin{minipage}{7.3cm}
\includegraphics[width=.97\textwidth]{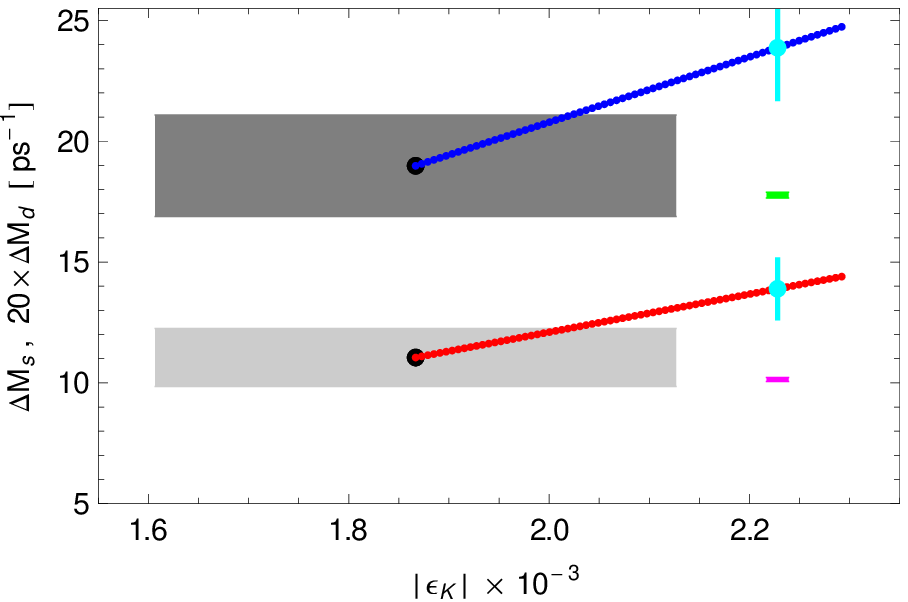}
\end{minipage}
\caption{{\it Left:} Correlation between $B_s\to\mu^+\mu^-$ and $B_d\to\mu^+\mu^-$ in MFV models. {\it Right:} Correlation between $\Delta M_{d,s}$ and $\varepsilon_K$ in CMFV models. Figures taken from \cite{Buras:2012ts} with kind permission of the authors.\label{fig1}}
\end{figure}

A generic prediction of MFV models is the correlation between the decays $B_s\to\mu^+\mu^-$ and $B_d\to\mu^+\mu^-$ shown in the left panel of Fig.\ \ref{fig1} \cite{Hurth:2008jc,Buras:2012ts}. While large deviations from the SM prediction in the $B_s$ channel have recently been excluded by LHCb and also Atlas and CMS, there is still a lot of room for NP effects in the $B_d$ channel. Any deviation from the straight linear correlation displayed by the red line would be a clear signal of new sources of flavor violation. This correlation, together with the data on $B_s\to\mu^+\mu^-$, also allows to derive a rough upper bound
\[
Br(B_d\to \mu^+\mu^-)<1.5\cdot10^{-10}
\]
valid in MFV models. An observation of $B_d\to \mu^+\mu^-$ above this limit would be an unambiguous sign of non-MFV interactions.

Further interesting correlations can be obtained by further restricting the framework. In the constrained MFV (CMFV) scenario in addition to the MFV hypothesis only the SM effective operators are assumed to be relevant at the electroweak scale \cite{Buras:2000dm,Buras:2003jf,Blanke:2006ig}. Consequently all new flavor violating effects can be described by the relevant CKM elements times a flavor universal factor -- in other words the {\it relative} size of the NP contribution is flavor universal. As an example, the right panel of Fig.\ \ref{fig1} shows the correlation between $\Delta M_{s,d}$ and $\varepsilon_K$ in CMFV models. Flavor universality predicts the straight line; furthermore in CMFV models only positive NP contributions are allowed \cite{Blanke:2006yh}. Consequently while CMFV is in principle able to solve the tension between $S_{\psi K_S}$ and $\varepsilon_K$ by enhancing the latter, simultaneously enhanced values for $\Delta M_{s,d}$ are predicted. This solution is clearly disfavored by the data and the recent lattice results, which prefer a slight suppression in $\Delta M_{s,d}$.

\section{Flavor correlations beyond MFV}

\begin{figure}
\begin{minipage}{7cm}
\includegraphics[width=.98\textwidth]{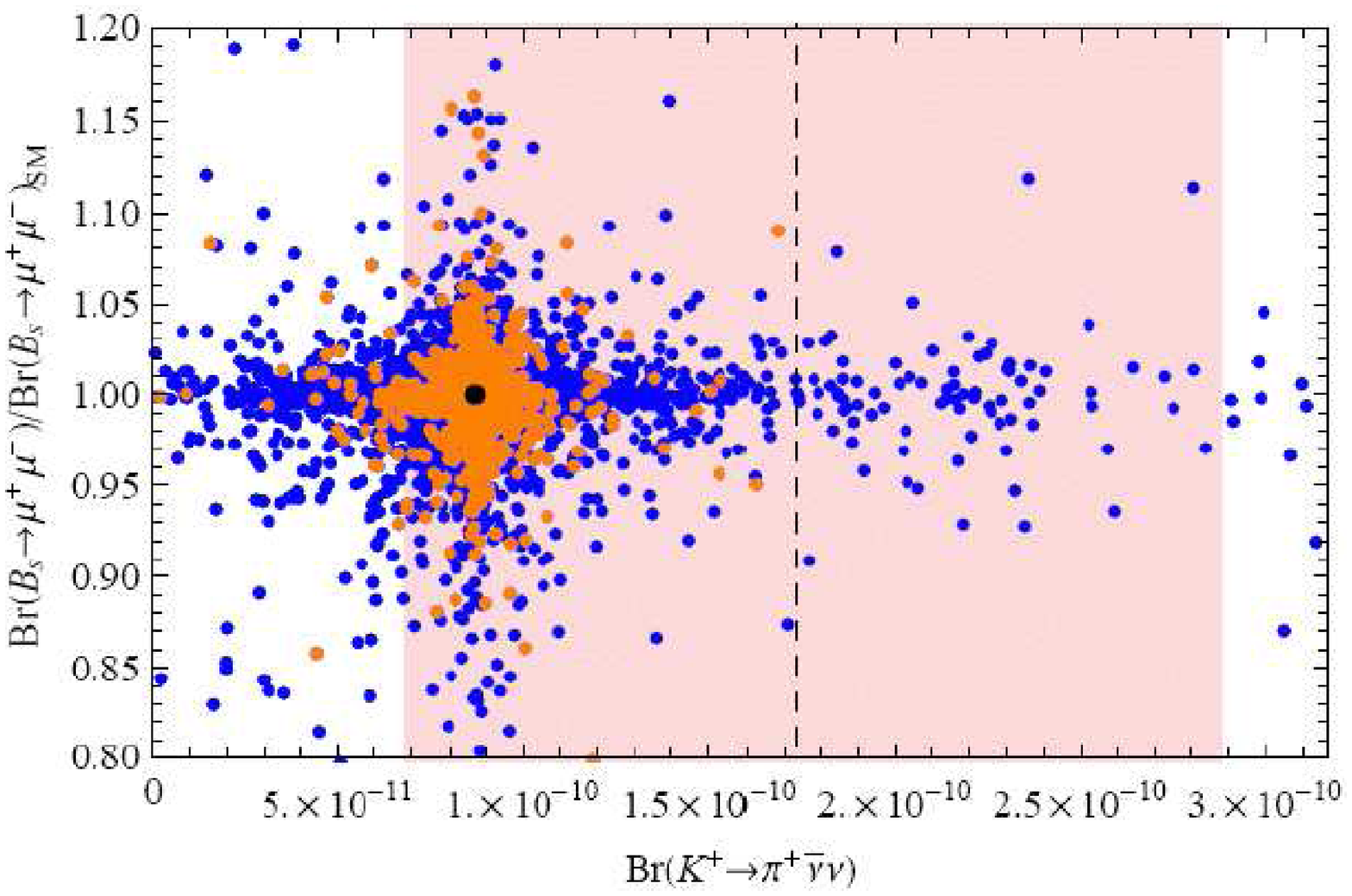}
\end{minipage}
\begin{minipage}{7.65cm}
\includegraphics[width=\textwidth]{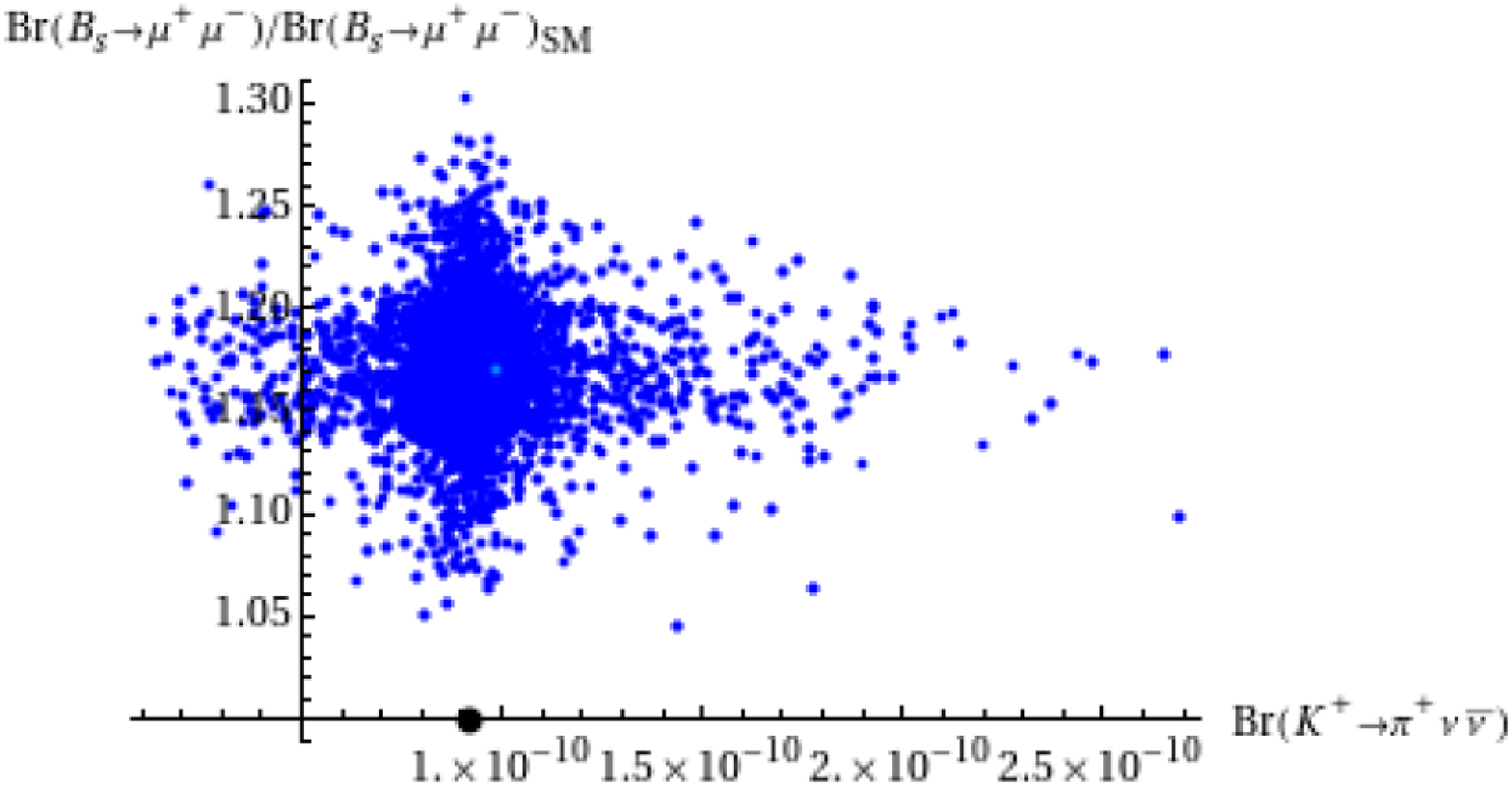}
\end{minipage}
\caption{Correlation between $K^+\to\pi^+\nu\bar\nu$ and $B_s\to\mu^+\mu^-$ in the RSc model \cite{Blanke:2008yr} and in the LHT model \cite{Blanke:2009am}.\label{fig2}}
\end{figure}

\begin{figure}
\includegraphics[width=.55\textwidth]{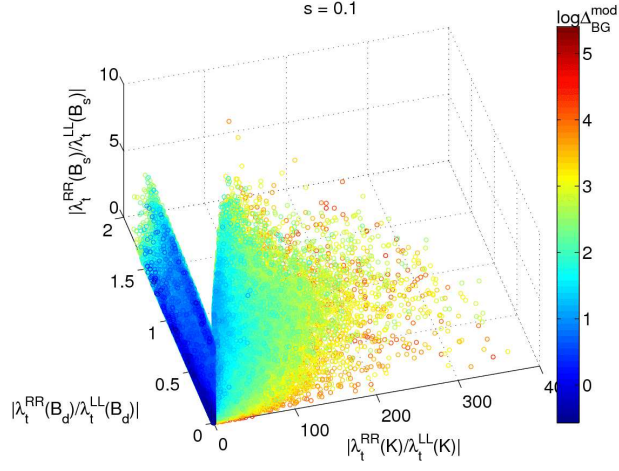}
\caption{Relative size of NP effects in the $K$, $B_d$ and $B_s$ systems in a general LR model \cite{Blanke:2011ry}.\label{fig3}}
\end{figure}

While MFV provides a viable and phenomenologically attractive framework, many NP scenarios predict non-MFV interactions. One of the most striking signatures of such interactions would certainly be a breakdown of universality in flavor violating NP effects. While the specific pattern of effects are model and parameter dependent, a general hierarchy of effects can be deduced from the structure of the CKM matrix, governing the size of flavor violating effects in the SM:
\[
\underbrace{V_{ts}^* V_{td}}_{K\;\mathrm{system}} \sim 5\cdot 10^{-4} \ll 
\underbrace{V_{tb}^* V_{td}}_{B_d\;\mathrm{ system}} \sim  10^{-2} < 
\underbrace{V_{tb}^* V_{ts}}_{B_s\;\mathrm{ system}} \sim  4\cdot 10^{-2}\,.
\]
Barring model-specific hierarchies in the NP flavor sector rare kaon decays, due to their strong CKM suppression in the SM, offer the largest NP sensitivity, while the effects in rare $B$ decays are generally expected to be much smaller. Such a pattern of NP effects can indeed be found e.\,g.\ in the Littlest Higgs model with T-parity (LHT) \cite{Blanke:2006eb,Blanke:2009am,Blanke:2012xy}, in the custodially protected Randall-Sundrum model (RSc) \cite{Blanke:2008yr,Albrecht:2009xr} or in a general left-right model (LR) \cite{Blanke:2011ry}, see Figs.\ \ref{fig2} and \ref{fig3}. Consequently even with SM-like effects at LHCb, large NP signatures can still be hoped for in rare kaon decays, such as the $K\to \pi\nu\bar\nu$ system.

\begin{figure}
\includegraphics[width=.55\textwidth]{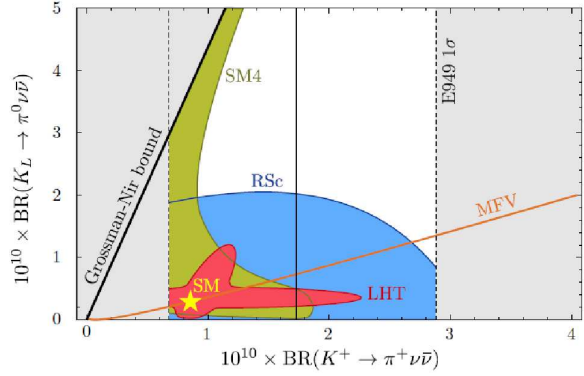}
\caption{Correlation between $K^+\to\pi^+\nu\bar\nu$ and $K_L\to\pi^0\nu\bar\nu$ in various NP models. Figure taken from \cite{Straub:2010ih} with kind permission of the author.\label{fig4}}
\end{figure}

The $K\to \pi\nu\bar\nu$ decays offer a particularly interesting place to look for physics beyond the SM, being highly suppressed in the SM and theoretically exceptionally clean (see \cite{Brod:2010hi} for the latest SM prediction). Beyond their unique NP discovery potential, the correlation between $K_L\to\pi^0\nu\bar\nu$ and $K^+\to\pi^+\nu\bar\nu$ offers valuable information on the NP model at work. In models like LHT where only the SM operators are present and $\Delta S=2$ and $\Delta S =1 $ transitions are strongly correlated, only two possible branches in the $K\to\pi\nu\bar\nu$ plane are allowed \cite{Blanke:2009pq} as the stringent constraint from $\varepsilon_K$ becomes effective here. On the other hand, in models like RS where $\Delta S=2$ is dominated by the chirally enhanced left-right operators, no correlation is visible, as seen in Fig.\ \ref{fig4}. In this sense the $K\to\pi\nu\bar\nu$ decays are able to shed light on the NP operator structure in $\Delta S = 2$ transitions.

\begin{figure}
\includegraphics[width=.55\textwidth]{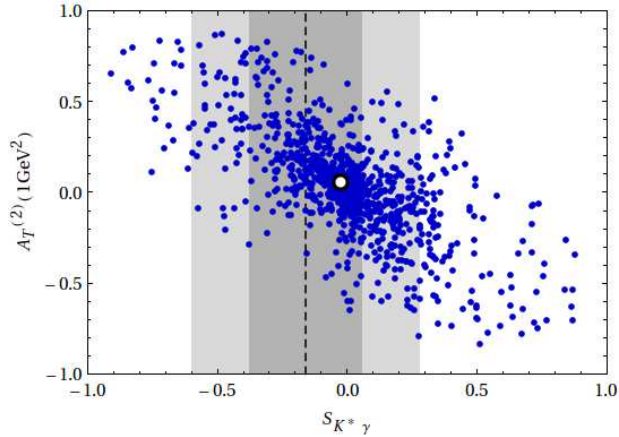}
\caption{Correlation between $S_{K^*\gamma}$ and $A_T^{(2)}$ in the RSc model \cite{Blanke:2012tv}.\label{fig5}}
\end{figure}

While rare $B$ decays cannot generally compete with rare $K$ decays in terms of NP discovery potential, they are able to provide valuable and complementary information on the NP flavor structure. Of particular interest are the $b\to s\gamma$ and $b\to s\mu^+\mu^-$ transitions, which are mediated by the dipole operators $C_7, C'_7$ in addition to the four fermion operators $C_9,C'_9$ and $C_{10},C'_{10}$. Although strong constraints exist from $Br(B\to X_s\gamma)$, $Br(B\to X_s\mu^+\mu^-)$, $Br(B_s\to\mu^+\mu^-)$ and $A_\mathrm{FB}(B\to K^*\mu^+\mu^-)$, the bounds on the chirality-flipped primed operators and CP asymmetries are still rather weak. Of particular interest are the time-dependent CP asymmetry in $B\to K^*\gamma$, and a number of angular observables in $B\to K^*\mu^+\mu^-$ \cite{Bobeth:2008ij,Altmannshofer:2008dz,Egede:2010zc,Altmannshofer:2011gn,Altmannshofer:2012ir}, allowing to disentangle the operator structure of the underlying NP. As an example we show in Fig.\ \ref{fig5} the correlation between the time-dependent CP asymmetry $S_{K^*\gamma}$ and the transverse asymmetry $A_T^{(2)}$, evaluated at $q^2=1\,\mathrm{GeV}$, in the RSc model \cite{Blanke:2012tv}. In that model NP contributions dominantly affect the chirality-flipped photon penguin operator $C'_7$ \cite{Agashe:2004ay,Blanke:2012tv}, yielding a very specific pattern of effects.

\section{Conclusions}

Correlations between flavor violating observables offer a powerful tool to disentangle the NP operator and flavor structure and therefore allow to discriminate among various NP scenarios. With the LHCb currently taking data and the next generation $B$ factories as well as dedicated kaon experiments starting operation in the foreseeable future, we can hope to soon shed some light on the underlying physics of flavor.

\begin{theacknowledgments}
This work is supported in part by the U.S. National Science Foundation
through grant PHY-0757868 and CAREER award PHY-0844667.
\end{theacknowledgments}



\bibliographystyle{aipproc}   

\bibliography{108_Blanke}

\IfFileExists{\jobname.bbl}{}
 {\typeout{}
  \typeout{******************************************}
  \typeout{** Please run "bibtex \jobname" to optain}
  \typeout{** the bibliography and then re-run LaTeX}
  \typeout{** twice to fix the references!}
  \typeout{******************************************}
  \typeout{}
 }

\end{document}